\documentclass[12pt]{iopart}
\usepackage{iopams}
\usepackage{graphicx}
\usepackage{braket}
\usepackage{dsfont}
\usepackage{color}
\usepackage{booktabs}
\usepackage[cdot,squaren, thinspace,thinqspace]{SIunits}
\usepackage{bm}
\usepackage{cite}

\definecolor{darkgreen}{rgb}{0,0.4,0}

\newcommand{\h}[1]{\hat{#1}}

\begin{document}

\title[]{Spatio-spectral characteristics of parametric down-conversion in waveguide arrays}

\author{R. Kruse$^{1}$, F. Katzschmann$^{1}$,  A. Christ$^{1}$,
  A. Schreiber$^{1,2}$,\\ S. Wilhelm$^{1}$, K. Laiho$^{1,3}$,
  A. G\'abris$^{4,5}$, C. S. Hamilton$^{4}$, I. Jex$^{4}$ and C. Silberhorn$^{1,2}$}
\address{$^{1}$Applied Physics, University of Paderborn,  Warburger Stra\ss e 100, 33098 Paderborn, Germany}
\address{$^{2}$Max Planck Institute for the Science of Light, G\"unther-Scharowsky Stra\ss e 1 - Building 24, 91058 Erlangen, Germany}
\address{$^3$University of Innsbruck, Institute for Experimental Physics/Photonics, Technikerstr. 25d, 6020 Innsbruck, Austria}
\address{$^{4}$Department of Physics, Faculty of Nuclear Sciences and Physical Engineering, Czech Technical University in Prague, B{\v r}ehov\'a 7, 115 19 Praha, Czech Republic.}
\address{$^{5}$Wigner Research Centre for Physics, Hungarian Academy of Sciences, H-1525 Budapest, P. O. Box 49, Hungary}

\ead{rkruse2@mail.upb.de \\
\date{\today}}

\begin{abstract}
High dimensional quantum states are of fundamental interest for quantum information processing. They give access to large Hilbert spaces and, in turn, enable the encoding of quantum information on multiple modes. One method to create such quantum states is parametric down-conversion (PDC) in waveguide arrays (WGAs) which allows for the creation of highly entangled photon pairs in controlled, easily accessible spatial modes, with unique spectral properties.

In this paper we examine both theoretically and experimentally the PDC process in a lithium niobate WGA. We measure the spatial and spectral properties of the emitted photon pairs, revealing correlations between spectral and spatial degrees of freedom of the created photons. Our measurements show that, in contrast to prior theoretical approaches, spectrally dependent coupling effects have to be taken into account in the theory of PDC in WGAs. To interpret the results, we developed a theoretical model specifically taking into account spectrally dependent coupling effects, which further enables us to explore the capabilities and limitations for engineering the spatial correlations of the generated quantum states.
\end{abstract}

\newpage

\section{Introduction}
The process of parametric down-conversion (PDC) \cite{burnham_observation_1970} provides a versatile resource for the creation and engineering of sophisticated quantum states of light. In particular, the creation of photon-pair states, entangled in various degrees of freedom, is of special interest for quantum information and quantum communication applications. To date several experiments have investigated the entanglement of these photon pairs in various degrees of freedom, for example using their spectral properties \cite{keller_theory_1997, grice_eliminating_2001, fedorov_anisotropically_2007, brida_characterization_2009, christ_probing_2010}, the orbital angular momentum of the photons \cite{mair_entanglement_2001, molina-terriza_twisted_2008, leach_quantum_2010}, transverse modes of a single waveguide \cite{mosley_direct_2009}, polarization \cite{kwiat_new_1995, levine_polarization-entangled_2011} and energy-time \cite{rarity_two-photon_1990, ou_observation_1990}.

Recently, PDC in a waveguide array (WGA) with a nonlinear optical response has been proposed as a resource for preparing highly entangled photon pairs into controllable and easily accessible spatial modes \cite{solntsev_spontaneous_2012, solntsev_photon-pair_2012,solntsev_simultaneous_2012}. In a nonlinear WGA several waveguides are embedded into a nonlinear material, such as lithium niobate \cite{iwanow_arrays_2005}, and are grouped so close to each other, that the created PDC photons are able to couple evanescently to the nearest-neighbour waveguides. This introduces, aside from the spectral degree of freedom present in the PDC process \cite{karpinski_experimental_2009, saleh_modal_2009, christ_spatial_2009}, a new spatial degree of freedom, where each waveguide channel represents an individual, easily accessible spatial mode. In general, the use of the spatial degree of freedom is a well-known tool for state engineering in bulk crystals \cite{uren_photon_2003}. PDC in WGAs, however, benefits from the bright and efficient photon-pair production in waveguides  \cite{serkland_squeezing_1995, tanzilli_highly_2001}, and offers several further advantages: these systems are not subjected to any in-coupling losses and thus maintain the process efficiency and quantum features. Further, a miniaturized, on-chip realization is also highly beneficial concerning the scalability, stability, adjustability and coherence of the system. Moreover, active optical devices can be implemented in the nonlinear substrate of lithium niobate, thus allowing the implementation of flexible, fast switching integrated optical components on the chip for controlling the properties of quantum light \cite{bonneau_fast_2012}.

The linear and nonlinear characteristics of WGAs have attracted enormous interest for the investigation of classical wave phenomena \cite{lederer_discrete_2008}. In the context of quantum optics these systems enable the implementation of quantum walks \cite{perets_realization_2008, bromberg_quantum_2009, owens_two-photon_2011}. In combination with non-classical light sources, WGAs have further been utilized for the study of bosonic and fermionic behavior \cite{peruzzo_quantum_2010, sansoni_two-particle_2012, crespi_anderson_2013}. WGAs have also become a resource for simulating the properties of other quantum systems. Utilizing classical light sources, WGAs with different types of architectures have been used to mimic the properties of photon-pair states \cite{grafe_biphoton_2012}, displaced Fock states \cite{keil_r.and_perez-leija_classical_2011}, and faithful quantum state transfer \cite{bellec_faithful_2012}.

Consequently, integrating WGAs in nonlinear materials offers a variety of possibilities for quantum optical experiments and a high potential for integrated quantum state generation and manipulation. For this purpose a detailed and accurate understanding of the process properties is necessary to enable precise quantum state engineering.

In this paper, we theoretically as well as experimentally, investigate PDC in a nonlinear WGA with special attention to the correlations between the spatial and spectral degrees of freedom. We explicitly take into account frequency dependent coupling effects and show how the frequencies of the created photon pairs are connected with their spatial properties. With this knowledge we then explore the preparation of specific spatial correlations between the created photon pairs.

The paper is structured into three main parts. In section \ref{sec:pdc_in_nonlinear_waveguide_arrays}, we introduce our theoretical PDC model including spectrally dependent coupling effects. In section \ref{sec:spatio-spectral_dynamics_in_waveguide_arrays} we investigate the occurring spatio-spectral correlations and introduce different ways to utilize them in order to create highly sophisticated quantum states. Finally, in section \ref{sec:experiment}, we perform a spectrally and spatially resolved measurement of the PDC emission in a periodically poled lithium niobate (PPLN) WGA. Our measurement results demonstrate the predicted spatio-spectral correlations and confirm the necessity to include spectrally dependent coupling effects into the theoretical treatment of PDC in a WGA. Furthermore our experimental investigations enable us to extract the phase-matching curve of the WGA from the acquired data, which is in very good agreement with the theoretical prediction.

\section{PDC in nonlinear WGAs}\label{sec:pdc_in_nonlinear_waveguide_arrays}
\begin{figure}[htb]
    \begin{center}
        \includegraphics[width = 0.5\textwidth]{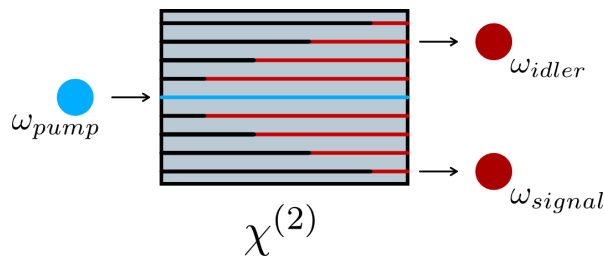}
    \end{center}
    \caption{Sketch of PDC in a WGA: a pump photon, in the illuminated waveguide channel decays via the $\chi^{(2)}$ nonlinearity, into a signal and idler photon pair which then couples from waveguide to waveguide until it exits the crystal.}
    \label{fig:pdc_scheme}
\end{figure}

During the PDC process a pump photon is converted inside a nonlinear crystal into a photon pair usually labelled signal and idler. In this paper we consider PDC in a WGA, where the generated photons can couple from waveguide to waveguide before they exit the crystal, while the pump is constrained to one waveguide, as schematically depicted in figure \ref{fig:pdc_scheme}.

\subsection{Electric fields in WGAs}
Following the theoretical treatment of type-I PDC in nonlinear WGAs by Solntsev et al. in \cite{solntsev_spontaneous_2012} we start with the mathematical description of the quantized electric fields propagating through the WGA and use coupled mode theory to solve the corresponding Maxwell's equations \cite{trompeter_discrete_2006, makris_method_2006, szameit_hexagonal_2006, szameit_light_2007}. The nearest neighbour coupling in the WGA induces a modified, discretized dispersion \cite{lederer_discrete_2008} altering the propagation vector \(\beta\) in the \(z\)-direction
\begin{eqnarray}
    \beta(\omega,k^\perp)=\beta^{(0)}(\omega) + 2C(\omega)\cos(k^\perp),
    \label{eq:propagation_constant_waveguide_array}
\end{eqnarray}
where \(\beta^{(0)}(\omega)\) describes the propagation vector in a single waveguide at frequency \(\omega\), given by \(\beta^{(0)}(\omega) = n^{(\mathrm{eff})}(\omega) \omega / c\), and \(k^\perp=k_x\cdot d\) is its normalized transverse momentum, with $d$ as the distance between the waveguides\footnote{A more detailed introduction to transverse momenta of particles and bandstructures may be found in \cite{lederer_discrete_2008} or \cite{kittel_introduction_2004}.}. The effective refractive index \(n^{(\mathrm{eff})}(\omega)\) characterizes the dispersion of a light field which propagates inside a single waveguide taking into account material and modal dispersion contributions \cite{christ_spatial_2009}. The term \( 2C(\omega)\cos(k^\perp)\) corresponds to the modification of dispersion due to the presence of the WGA. Hereby the impact of the WGA on the dispersion is given by the coupling parameter \(C(\omega)\), which is firstly dependent on the distance between the individual waveguide channels and secondly impacted by the frequencies of the propagating fields (see \ref{app:wavelength_dependency_of_the_coupling_parameter}). The signal and idler fields $\h{E}_{n}^{(+)}(z,t)$ in the $n$th waveguide can be expressed in the form
\begin{eqnarray}
\fl \qquad \h{E}_{n}^{(+)}(z,t)=\h{E}_{n}^{(-)\dagger}(z,t) =
 B\int \limits_{-\pi}^{\pi} \rmd k^\perp  \int \limits_{-\infty}^{\infty} \rmd \omega \  \rme^{i k^\perp n} \rme^{i[\beta(\omega,k^\perp)z-\omega t]} \h{a}(\omega,k^\perp),
\label{eq:electric_field_waveguide_array}
\end{eqnarray}
where \(B\) collects all constants \cite{blow_continuum_1990}. The pump field driving the PDC process is considered as a bright, undepleted optical beam, which allows us to treat it classically. Moreover, the pump does not couple to neighbouring channels, because it resides at wavelengths far below signal and idler. In this regime the waveguide mode size is smaller, and thus, the mode overlap governing the coupling parameter can be approximated as zero. Therefore the influence of the WGA on the propagation vector can be neglected (see \ref{app:wavelength_dependency_of_the_coupling_parameter}). Thus, the pump ($p$) field in the $n$th waveguide is described as
\begin{eqnarray}
     \fl \qquad E_{p,n}^{(+)}(z,t)=E_{p,n}^{(-)*}(z,t) &=&  \int \limits_{-\infty}^{\infty} \rmd\omega_p{ \hspace{1ex} A(n)\alpha(\omega_p)\rme^{i \left (\beta^{(0)}(\omega_p)z-\omega_pt \right)}} \nonumber \\
    &=& \int \limits_{-\pi}^\pi \rmd k_p^\perp{ \int \limits_{-\infty}^{\infty} \rmd\omega_p{ \hspace{1ex}\alpha(\omega_p)\tilde{A}(k_p^\perp)\rme^{i k_p^\perp n}}\rme^{i \left (\beta^{(0)}(\omega_p)z-\omega_pt \right)}}, 
    \label{eq:pump_field_waveguide_array}
\end{eqnarray}
where \(\alpha(\omega_p)\) is the spectral shape of the pump beam. The spatial illumination pattern of the pump \(A(n)\) in the $n^{\mathrm{th}}$ waveguide is connected to its Bloch mode distribution \(\tilde{A}(k_p^\perp)\) via a Fourier transformation $\tilde{A}(k_p^\perp)=\frac{1}{2\pi}\sum_n{A(n) \rme^{-i k_p^\perp n}}$.

\subsection{The PDC state}\label{sec:the_pdc_state}
Using the electric field definitions in \eref{eq:electric_field_waveguide_array} and \eref{eq:pump_field_waveguide_array}, we are able to express the effective Hamiltonian of the PDC process in a nonlinear WGA by
\begin{eqnarray}
\h{H}_{\mathrm{PDC}}(t)&=&\frac{\epsilon_0}{2} \int \limits_{-L}^{0}
\hspace{-1ex} \rmd z \sum_n{\chi^{(2)}\left[E_{p,n}^{(+)}(z,t) \h{E}_{n}^{(-)}(z,t)  \h{E}_{n}^{(-)} (z,t) +h.c.\right]},
\label{eq:interaction}
\end{eqnarray}
where $\epsilon_{0}$ denotes the electric constant, $\chi^{(2)}$ is the nonlinearity of the material and \(L\) the length of the WGA with an infinite number of channels \(n\).

The nonlinear interaction inside the medium is weak, which enables us to calculate the output state using first-order perturbation theory
\begin{eqnarray}
    \ket{\psi} \approx \ket{0} -\frac{i}{\hbar}\int
    \limits_{-\infty}^\infty \rmd t \, \h{H}_{\mathrm{PDC}}(t) \ket{0}.
 \label{eq:pdc_state_schroedinger_perturbation}
\end{eqnarray}
%
We have extended the integration boundaries to plus and minus infinity, since we regard the PDC state long after the interaction in the crystal \cite{grice_spectral_1997}.
We further post-select on the detection of photon pairs, which enables us to drop the vacuum contribution appearing in \eref{eq:pdc_state_schroedinger_perturbation} and renormalize the state accordingly. A straightforward calculation delivers the expression for the two-photon PDC state emerging from the WGA as
\begin{eqnarray}
    \fl \qquad \ket{\psi} = \frac{1}{\sqrt{\mathcal{N}}}   \int \limits_{-\infty}^{\infty} \hspace{-2mm} \rmd\omega_s  \hspace{-2mm} \int \limits_{-\infty}^{\infty} \hspace{-2mm} \rmd\omega_i \hspace{-1mm}  \int \limits_{-\pi}^{\pi} \hspace{-2mm} \rmd k^\perp_{s} \hspace{-0mm} \int \limits_{-\pi}^{\pi}  \hspace{-2mm} \rmd k^\perp_{i}   \ f(\omega_s, \omega_{i}, k_s^\perp,k_i^\perp)  \hat{a}^\dagger(\omega_s,k_s^\perp)\hat{a}^\dagger(\omega_i,k_i^\perp) |0\rangle ,
\label{eq:pdc_state}
\end{eqnarray}
i.e., two photons are created into a superposition of spectral and spatial modes, where $1/\sqrt{\mathcal{N}}$ is the normalization constant. The spatial and spectral structure of the created photon pair is determined by the form of the joint spatio-spectral amplitude \(f(\omega_s, \omega_{i}, k_s^\perp,k_i^\perp)\) of the generated signal ($s$) and idler ($i$) photons. It assumes the form
\begin{eqnarray}
\nonumber
\fl \qquad f(\omega_s, \omega_{i}, k_s^\perp,k_i^\perp) &=&
\alpha(\omega_s+\omega_i) \tilde{A}(k_s^\perp+k_i^\perp) \\
\nonumber
&\times& \underbrace{\textrm{sinc}\left[\frac{L}{2} \Delta\beta(\omega_s,\omega_i, k^\perp_s,k^\perp_i) \right]}_{\Phi(\omega_s, \omega_i, k_s^\perp, k_i^\perp)} \underbrace{\exp\left[-i \Delta \beta(\omega_s,\omega_i, k^\perp_s,k^\perp_i) \frac{L}{2}\right]}_{\varphi(\omega_s, \omega_i, k_s^\perp, k_i^\perp)}\\
&=& \alpha(\omega_s+\omega_i) \tilde{A}(k_s^\perp+k_i^\perp) \Phi(\omega_s, \omega_i, k_s^\perp, k_i^\perp) \varphi(\omega_s, \omega_i, k_s^\perp, k_i^\perp),
\label{eq:pdc_state_spatio-spectral_amplitude}
\end{eqnarray}
in which  the phase-mismatch \(\Delta \beta (\omega_{s}, \omega_{i}, k^\perp_s,k^\perp_i) \) is  defined as
\begin{eqnarray}
    \Delta \beta(\omega_{s}, \omega_{i}, k^\perp_s,k^\perp_i) = \beta^{(0)}_p(\omega_s+\omega_i)-\beta(\omega_s,k_s^\perp)-\beta(\omega_i, k_i^\perp).
    \label{eq:delta_beta}
\end{eqnarray}
The four terms determining the properties of the two-photon state in \eref{eq:pdc_state_spatio-spectral_amplitude} can be attributed to the following sources. The spectral and spatial properties of the pump field determine the functions $\alpha(\omega_s+\omega_i)$ and $\tilde{A}(k_s^\perp+k_i^\perp)$ respectively. The phase-matching function $\Phi(\omega_s,\omega_i,k^\perp_s,k^\perp_i)$ and the phase factor \(\varphi(\omega_s, \omega_i, k_s^\perp, k_i^\perp)\) are given by the nonlinear and linear properties of the photonic structure, i.e. the dispersion in the medium, periodic poling and waveguide spacing.

\section{Spatio-spectral properties of PDC in WGAs}\label{sec:spatio-spectral_dynamics_in_waveguide_arrays}

\subsection{Origin of the spatio-spectral correlations between the created photon pairs}\label{sec:the_phasematching_function}

The key for understanding the spatio-spectral properties of the generated PDC state lies in the phase-matching function \(\Phi(\omega_s, \omega_i, k_s^\perp, k_i^\perp)\). Significant down-conversion rates require nearly perfect phase-matching, satisfying \(\Delta\beta(\omega_s,\omega_i, k^\perp_s,k^\perp_i) \approx 0\). This constraint induces spatio-spectral correlations between the frequencies \( (\omega_s, \omega_i) \) and \(k^\perp\)-components \( (k_s^\perp, k_i^\perp) \) of the generated signal-idler photon pairs.

In order to understand its exact nature we split the phase-mismatch \(\Delta\beta(\omega_s,\omega_i, k^\perp_s,k^\perp_i)\) in equation \eref{eq:delta_beta}, into its contributions from the single, isolated waveguide \(\Delta \beta_\omega\) and the dispersion effects from the WGA \(\Delta \beta_A\)
\begin{eqnarray}
    \Delta\beta(\omega_s,\omega_i, k^\perp_s,k^\perp_i) = \Delta \beta_\omega(\omega_s, \omega_i) + \Delta \beta_A(k_s^\perp, k_i^\perp, \omega_s, \omega_i).
    \label{eq:phase-mismatch}
\end{eqnarray}
The first term describes the spectral phase-mismatch, known from PDC in a single isolated waveguide,
\begin{eqnarray}
    \Delta \beta_\omega(\omega_s, \omega_i) = \beta^{(0)}_p(\omega_s + \omega_i) - \beta^{(0)}(\omega_s) - \beta^{(0)}(\omega_i),
    \label{eq:spectral_phase_mismatch}
\end{eqnarray}
which stems mostly from the dispersion of the material with a contribution from the modal properties of the waveguides \cite{christ_spatial_2009}. The second term is the phase-mismatch induced by the dispersion of the WGA,
\begin{eqnarray}
    \Delta \beta_A(k_s^\perp, k_i^\perp, \omega_s, \omega_i) =  - 2 C(\omega_s) \cos(k_s^\perp) - 2 C(\omega_i) \cos(k_i^\perp),
    \label{eq:spatial_phase_mismatch}
\end{eqnarray}
which depends on the transverse \(k^\perp\)-components of the generated photons \( \left(k_s^\perp, k_i^\perp \right) \) and the frequency dependent coupling parameter \(C(\omega)\).

\begin{figure}[htb]
    \begin{center}
        \includegraphics[width=0.8\textwidth]{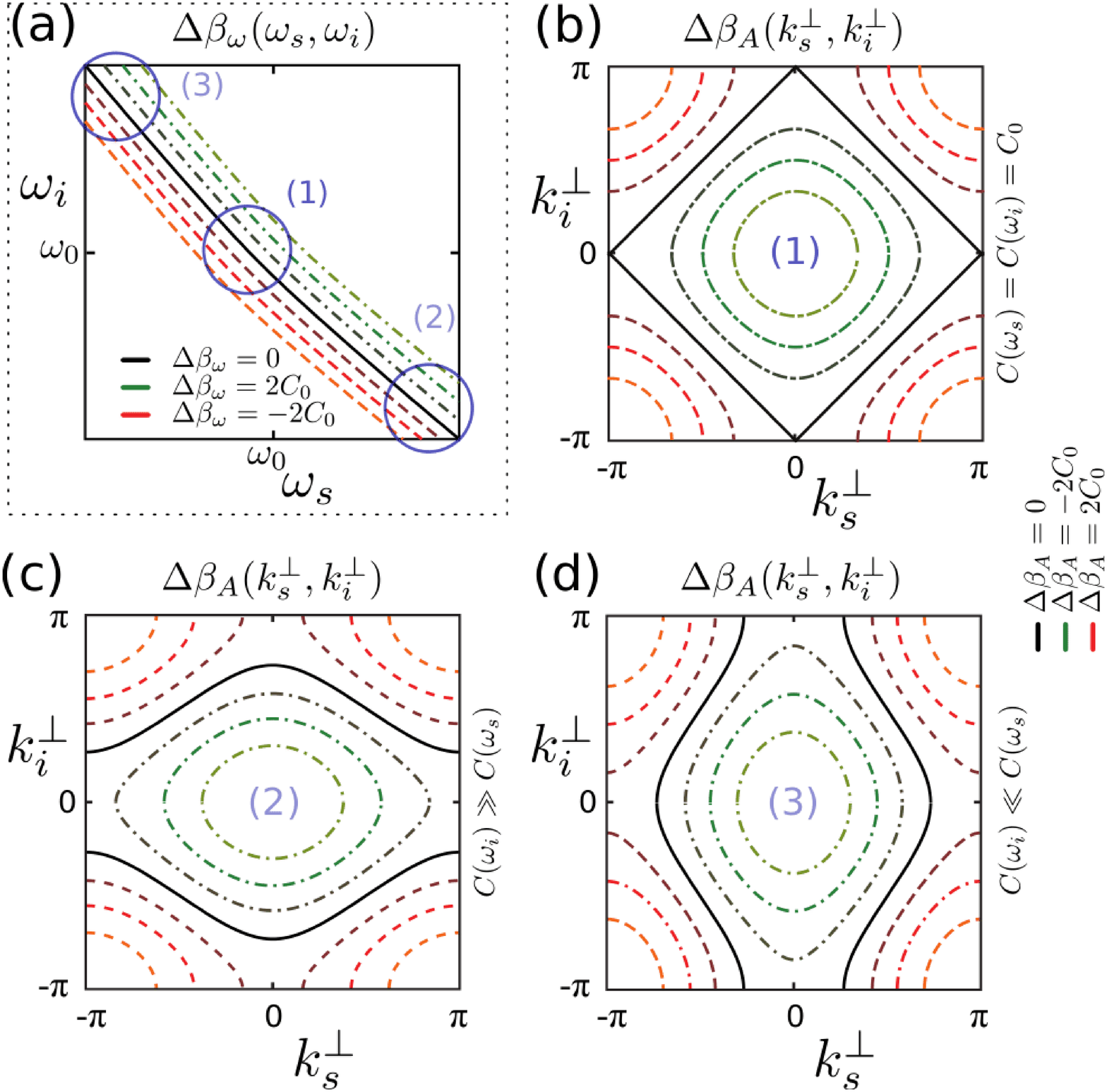}
    \end{center}
    \caption{Phase-mismatch \( \Delta \beta \) in the spectral and spatial domain. (a) In the spectral domain the phase-mismatch \(\Delta \beta_\omega(\omega_s, \omega_i)\) consists of curved lines oriented along the $-45^\circ$ axis in frequency space. (b) \(\Delta \beta_A(k_s^\perp, k_i^\perp)\) is defined by \(\Delta \beta_A(k_s^\perp, k_i^\perp, \omega_s, \omega_i)\) for frequencies \(\omega_s, \omega_i\) as depicted in (a). For degenerate frequencies \(\omega_s \approx \omega_i\) implying identical coupling parameters for signal and idler, \(\Delta \beta_A(k_s^\perp, k_i^\perp)\) is given by circular shapes around the middle or the edges of the Brillouin zone with corresponding color coding. If the signal and idler frequencies are distinct, the \(\Delta \beta_A(k_s^\perp, k_i^\perp)\) patterns get stretched from their symmetric shape in (b) into the forms depicted in (c) \(C(\omega_i) \gg C(\omega_s)\) and (d) \( C(\omega_i) \ll C(\omega_s)\).}
    \label{fig:phase-mismatch}
\end{figure}

In figure \ref{fig:phase-mismatch}(a) we sketch the contours of the spectral phase-mismatch \(\Delta\beta_\omega\) defined in \eref{eq:spectral_phase_mismatch} for various values. Due to dispersion effects of the material, this yields curved lines aligned along the $-45^\circ$-axis, for constant $\Delta\beta_\omega$ in \( (\omega_s, \omega_i) \)-space. In the case of negative spectral phase-mismatch (\(\Delta \beta_\omega < 0\)) the contours (red lines) for different values of \(\Delta \beta_\omega\) run below the central contour, colour-coded in black, (\(\Delta \beta_\omega = 0\)), while for positive spectral phase-mismatch (\(\Delta \beta_\omega > 0\)) the contours run above (green lines).

In order to achieve phase-matching in the WGA, i.e \(\Delta \beta = 0\), the spectral phase-mismatch \(\Delta \beta_\omega\) has to be compensated by a spatial phase-mismatch of \(\Delta \beta_A = -\Delta \beta_\omega \) with opposite sign to satisfy \(\Delta\beta(\omega_s,\omega_i, k^\perp_s,k^\perp_i) \approx 0\). The pairing of these values, $\Delta\beta_\omega$ and $\Delta\beta_A$ creates the spatio-spectral correlations between the PDC photons. The different contours for which the different values of \(\Delta\beta_\omega\) and \(\Delta\beta_A\) yield perfect phase-matching are colour-coded in figure \ref{fig:phase-mismatch}(a)-(d).

To investigate the induced correlations we choose three different scenarios for \(\Delta \beta_A\): \(\omega_s \gg \omega_i\), \(\omega_s \approx \omega_i\) and \(\omega_s \ll \omega_i\), as shown in figures \ref{fig:phase-mismatch}(b)-(d).

In figure \ref{fig:phase-mismatch}(b) we show the $(k_s^\perp,k_i^\perp)$ combinations yielding a constant $\Delta\beta_A$ at the degeneracy point (\(\omega_s \approx \omega_i\)). A numerical simulation illustrating this scenario can be found in \ref{app:numerical_simulation_of_the_waveguide_array}. In this regime both coupling parameters are identical and, in the considered frequency range, constant: \(C(\omega_s) = C(\omega_i) = C_0\).  In order to understand the depicted shapes we regard three different cases, where (i) \(\Delta \beta_A = 0\), (ii) \(\Delta \beta_A = -2C_0\) and (iii) \(\Delta \beta_A = 2C_0\):

\begin{description}
    \item[Scenario (i):] If \(\Delta \beta_A = 0\) the contributions from \(k_s^\perp\) and \(k_i^\perp\) in \eref{eq:spatial_phase_mismatch} cancel each other. This means that if the transverse momentum of the signal photon is in the center of the Brillouin zone\footnote{The Brillouin zone is the elemental cell of a lattice in Fourier space. Due to its periodic nature all momenta can be expressed in this part of the Fourier space\cite{kittel_introduction_2004}.} (\(k_s^\perp = 0\)) the momentum of the idler has to be at the edge (\(k_i^\perp = \pm \pi\)) or vice versa. This results in a rectangular shape in ($k^{\perp}_{s}, k^{\perp}_{i}$)-space, depicted in figure \ref{fig:phase-mismatch}(b).
    
    \item[Scenario (ii):] If $\Delta \beta_A = -2C_0$,  the correlations between $k_s^\perp$ and $k_i^\perp$ form a circle-like shape in the  ($k^\perp_{s}, k^\perp_{i}$)-space, as depicted by the green line in figure \ref{fig:phase-mismatch}(b). If either signal or idler is at the center of the Brillouin zone (\(k_s^\perp\) or \(k_i^\perp = 0\)) than the other one has to have a $k^\perp$-component with a value of $\pm\frac{\pi}{2}$ in order to reach the required phase-mismatch.

    \item[Scenario (iii):] If $ \Delta\beta_{A}= 2C_0$, the solutions are located at the corners of the ($k^\perp_{s}, k^\perp_{i}$)-space, as visualized by the red line in figure \ref{fig:phase-mismatch}(b). In order to reach the required negative phase-mismatch the transverse momentum of scenario (ii) has to be shifted from the center to the edges of the Brillouin zone.
\end{description}

In figure \ref{fig:phase-mismatch}(c) we depict the case \( C(\omega_i) \gg C(\omega_s)\) in a frequency range where both coupling values are constant (\(C(\omega_s) = C_s,\,\, C(\omega_i) = C_i\)). The different coupling values lead to stretched contours, in comparison to figure \ref{fig:phase-mismatch}(b). Finally, in figure \ref{fig:phase-mismatch}(d) we depict \( C(\omega_i) \ll C(\omega_s)\), which is a flipped version of figure \ref{fig:phase-mismatch}(c), due to the fact that we regard type-I PDC, where signal and idler are interchangeable. Here the different coupling constants are selected such, that the difference is strong enough to impact the spatial correlations.

In conclusion our investigations reveal that the origins of the spatio-spectral correlations of PDC in WGA, are located in the phase-matching condition $\Delta\beta(\omega_s,\omega_i,k_s^\perp,k_i^\perp) \approx 0$. Every spatial correlation pattern, as depicted in figure \ref{fig:phase-mismatch}(b)-(d) has a fixed spatial phase-mismatch $\Delta\beta_A$, which must be compensated by an appropriate spectral phase-mismatch $\Delta\beta_\omega$, satisfying $\Delta\beta_A=-\Delta\beta_\omega$. Next, we will illustrate, how to utilize this spatio-spectral correlation in order to generate sophisticated spatial correlations between the generated signal and idler photons.

\subsection{Engineering spatial correlations via pump shaping}\label{sec:engineering_spatial_correlations_via_pump_shaping}

In section \ref{sec:the_phasematching_function} we have shown that the spatial and spectral correlations in the phase-mismatch \(\Delta \beta\) have a significant influence on the emitted photon pairs. Driving the PDC process in the WGA with an adapted spatio-spectral pump shape enables us to engineer a variety of different spatial correlations between the generated photon pairs. The emerging joint spatio-spectral amplitude is given by (\ref{eq:pdc_state_spatio-spectral_amplitude})
\begin{eqnarray}
   \fl \qquad f(\omega_s, \omega_i, k_s^\perp, k_i^\perp) = \alpha(\omega_s + \omega_i) \tilde{A}(k_s^\perp + k_i^\perp) \Phi(\omega_s, \omega_i, k_s^\perp, k_i^\perp) \varphi(\omega_s, \omega_i, k_s^\perp, k_i^\perp).
    \label{eq:joint_spatio_spectral_amplitude_2}
\end{eqnarray}
This means we can regard \(\alpha(\omega_s + \omega_i) \tilde{A}(k_s^\perp + k_i^\perp)\) as a function which, via multiplication, selects a subset of the spatio-spectral correlations \(\Phi(\omega_s, \omega_i, k_s^\perp, k_i^\perp)\) inherent in the WGA. The phase \(\varphi(\omega_s, \omega_i, k_s^\perp, k_i^\perp)\) is neglected in the discussion, since it does not influence the $k^\perp$-space correlations, while only refining the correlation patterns in real space.

\begin{figure}[htb]
    \begin{center}
        \includegraphics[width=0.88\textwidth]{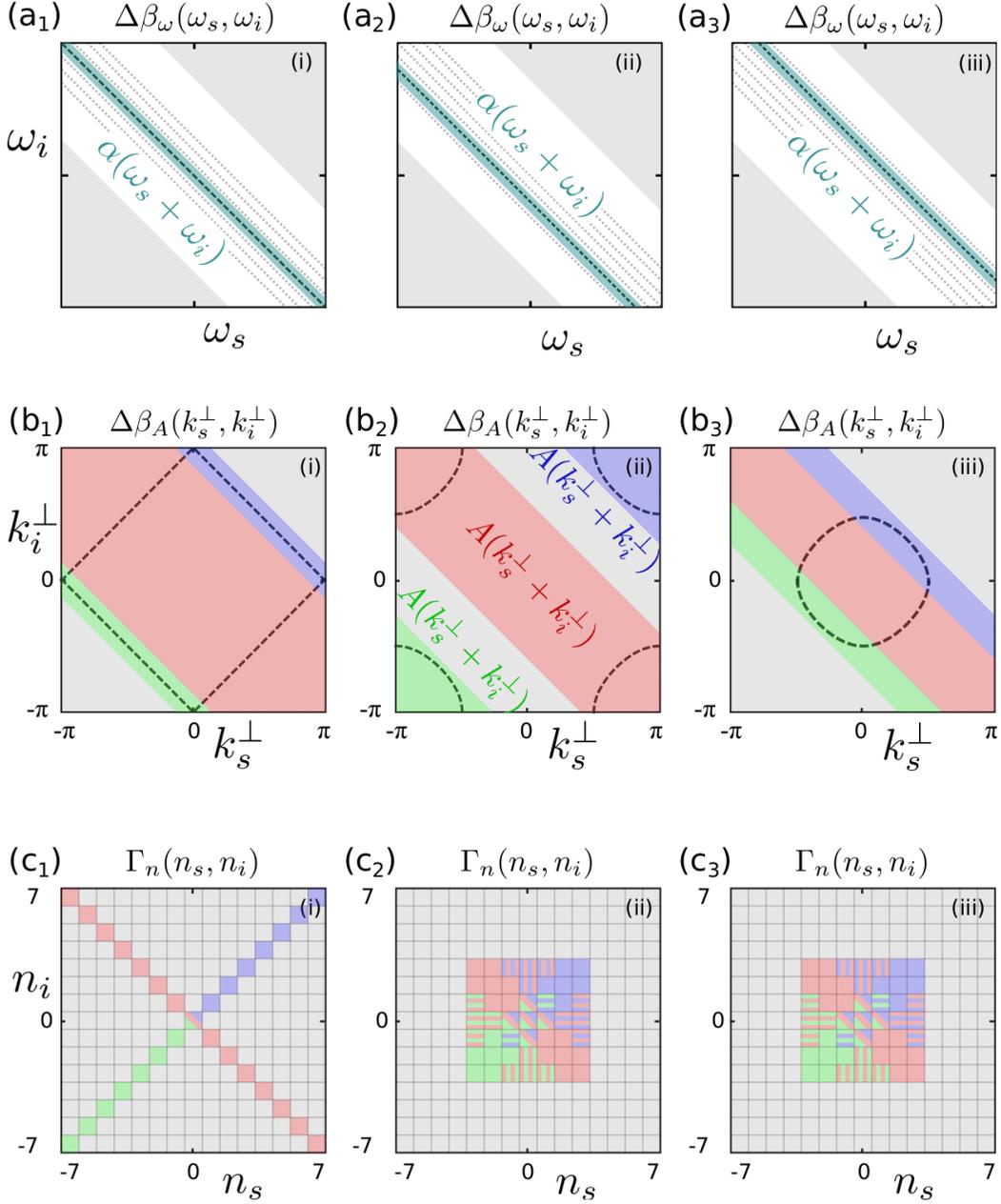}
    \end{center}
    \caption{In (a\(_1\))-(a\(_3\)) we show, how different spectral pump shapes (turquoise lines) \(\alpha(\omega_s + \omega_i)\) are able to select an individual frequency mismatch \(\Delta \beta_\omega\) (dashed grey lines). This enables us to excite photon pairs with a variety of \(k^\perp\)-correlations as depicted by the dashed black lines in (b\(_1\))-(b\(_3\)). These can further be modified by adapting the spatial pump shape \(\tilde{A}(k_s^\perp + k_i^\perp)\), by displacing the rectangular spatial pump shape in $k^\perp$-space, as colour-coded with green, red and blue in (b\(_1\))-(b\(_3\)). The corresponding photon-pair correlations in real space are depicted by red, green and blue squares in (c\(_1\))-(c\(_3\)). Due to the finite width of the spatial pump shape, the real space correlations will also have a finite width in $+45^\circ$-direction. The grey shaded areas in (a)-(c) indicate regions with no PDC signal. \ref{app:numerical_simulation_of_the_waveguide_array} gives a numerical simulation without spatial pump shaping.}
    \label{fig:reshaping_spatial_spectral}
\end{figure}

For degenerate signal and idler frequencies, the whole process is sketched, from top to bottom, in figure \ref{fig:reshaping_spatial_spectral}. This corresponds to the configuration presented in figure \ref{fig:phase-mismatch}(b). In this special case the spectral phase-mismatch \(\Delta \beta_\omega\) is aligned along the \(-45^\circ\)-axis in frequency space (dashed black lines in figure \ref{fig:reshaping_spatial_spectral} (a\(_1\))-(a\(_3\))).

Choosing a specific pump frequency (\(\omega_p = \omega_s + \omega_i\)) defines a line in frequency space, aligned along the \(-45^\circ\)-axis (torquoise shaded area in figure \ref{fig:reshaping_spatial_spectral} (a\(_1\))-(a\(_3\))) and consequently selects a specific phase-mismatch \(\Delta \beta_\omega\)\footnote{In this configuration the impacts of the curvature of \(\Delta \beta_\omega\) are negligible.}. Due to the spatio-spectral coupling, discussed in section \ref{sec:the_phasematching_function}, this in turn demands that all created photons satisfy \(\Delta \beta_A = - \Delta \beta_\omega\) in the spatial domain. The resulting \(k^\perp\)-correlations in the created photon pair for a flat spatial pump shape $\tilde{A}(k_s^\perp+k_i^\perp)=1$\footnote{This corresponds to pumping in a single waveguide.} are depicted by the dashed black lines in figure \ref{fig:reshaping_spatial_spectral}(b\(_1\))-(b\(_3\)), i.e. a rectangle in figure \ref{fig:reshaping_spatial_spectral}(b\(_1\)) for pumping at the central frequency, outward circles in figure \ref{fig:reshaping_spatial_spectral}(b\(_2\)) in the case of pumping below the central frequency and a circular shape in figure \ref{fig:reshaping_spatial_spectral}(b\(_3\)) for pumping above the central frequency. The corresponding correlations in real space, obtained via Fourier transformation, are depicted by the coloured squares in figure \ref{fig:reshaping_spatial_spectral}(c\(_1\))-(c\(_3\)), where each square represents a different signal-idler output channel combination. Here \(\Gamma_n(n_s, n_i)\) is the correlation function for measuring two photons at positions \( (n_s, n_i) \), as defined in \ref{app:appendix_correlation_function}. They vary from an X-shape in figure \ref{fig:reshaping_spatial_spectral}(c\(_1\)) to the square-like\footnote{For higher spectral phase-mismatch the shapes are of a circular form.} shapes sketched in figure \ref{fig:reshaping_spatial_spectral}(c\(_2\))-(c\(_3\)).

We can gain an enhanced control by adapting the spatial illumination of the WGA to modify \(\tilde{A}(k_s^\perp + k_i^\perp)\), i.e. we change from pumping in a single channel (\(\tilde{A}(k_s^\perp + k_i^\perp) = 1\)) to pumping in different channels simultaneously. Here, we choose a rectangularly formed spatial pump shape in $k^\perp$-space, which offers the advantage that we can finely select the spatial correlations. The spatial illumination pattern of the WGA to reach this shape in $k^\perp$-space is highly intricate, such that an experimental realization becomes more involved.

Three different scenarios are colour-coded in figure \ref{fig:reshaping_spatial_spectral}(b\(_1\))-(b\(_3\)) by the red, green and blue shaded areas, where we have displaced the rectangular spatial pump shape in $k^\perp$-space. Each consists of a different range of \(k_p^\perp\)-components specifically chosen to excite distinct correlations in real space. This simultaneous pumping in different channels enables us to individually select various parts of the phase-mismatch \(\Delta \beta_A\), as given by colour-coded areas, only constrained by the fact that the function \(\tilde{A}(k_s^\perp + k_i^\perp)\) has to be aligned along the \(-45^\circ\)-axis in \(k^\perp\)-space. This enables us, for example, to engineer photon bunching and antibunching effects\footnote{Due to the Fourier limitation, the finite width of the spatial pump will lead to finite widths in the real space correlations as well.}, as already discussed in \cite{solntsev_spontaneous_2012}, but it is also evident that steering the created photon pairs to the right or to the left is possible as well.

\subsection{Phase engineering of $\tilde{A}(k_s^\perp+k_i^\perp)$}

In the previous section we have concentrated on the influence of real valued pump shapes $\tilde{A}(k_s^\perp+k_i^\perp)$ on the correlation functions in both $k^\perp$- and real space. However, we are also able to engineer the signal-idler correlations by tuning the phase of the spatial pump distribution $\tilde{A}(k_s^\perp+k_i^\perp)$.
For this discussion, we restrict ourselves to the scenario discussed in section \ref{sec:engineering_spatial_correlations_via_pump_shaping} figure \ref{fig:reshaping_spatial_spectral}(a\(_1\))-(c\(_1\)) with a rectangular pump corresponding to an $\tilde{A}(k_s^\perp+k_i^\perp)$ selecting the red shaded area.

\begin{figure}
\begin{center}
\includegraphics[width=.9\textwidth]{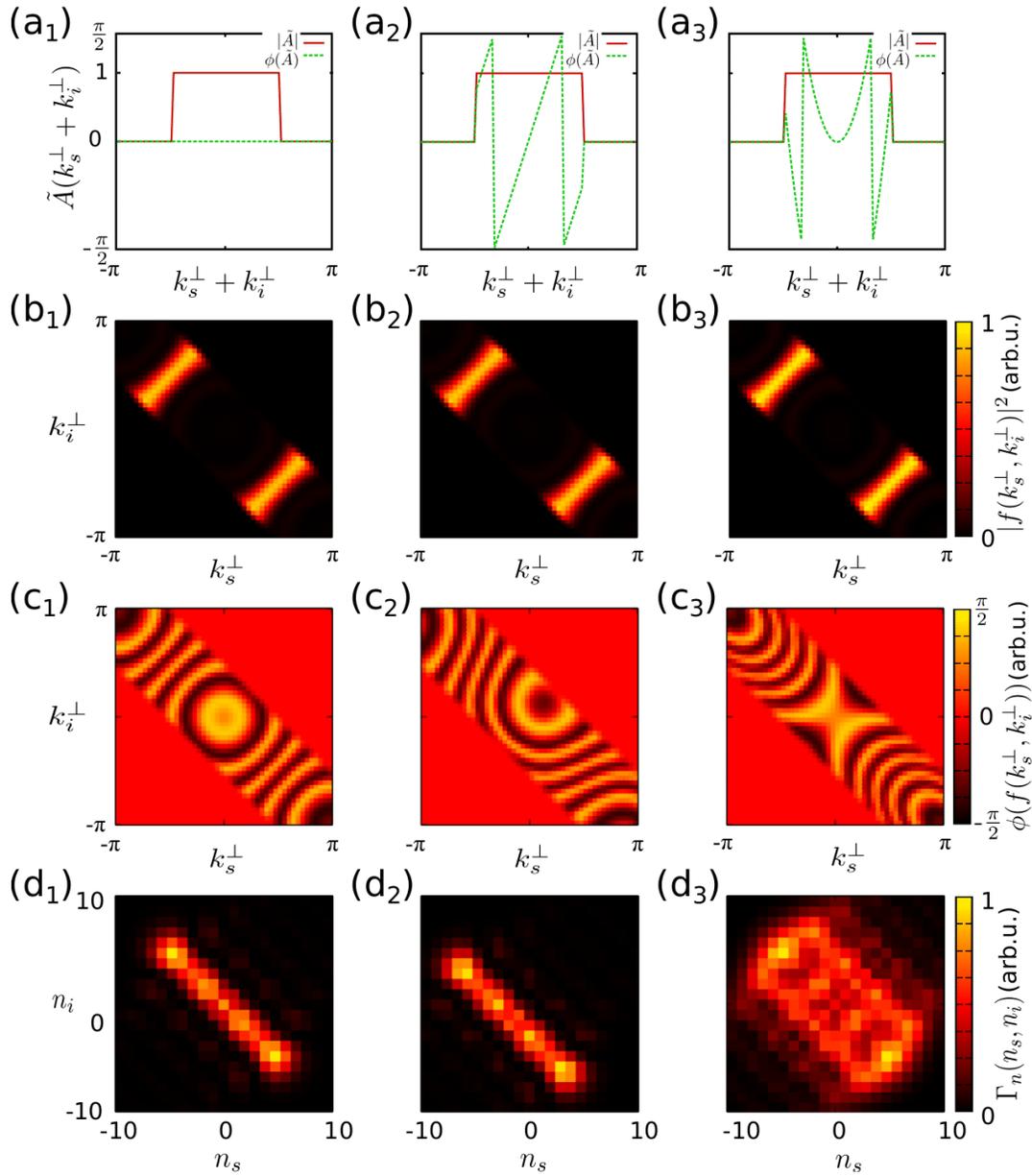}
\end{center}
\caption{Introducing different phase distributions in the spatial pump shape as indicated in (a$_1$)-(a$_3$) changes the real space correlations (d$_1$)-(d$_3$). Since the absolute value of the spatial pump shape does not vary, the $k^\perp$-space correlations in (b$_1$)-(b$_3$) remain unchanged, while the phase correlations in (c$_1$)-(c$_3$) show the influence of the pump phase. Linear phases introduce a shift of the real space correlations, without changing the internal shape of the correlation function, while quadratic phases introduce a "stretching" of the internal structure.}
\label{fig:phases}
\end{figure}

The applied pump shapes are given in figures \ref{fig:phases}(a$_1$)-(a$_3$), where red denotes the absolute value and green the phase of the function $\tilde{A}(k_s^\perp+k_i^\perp)$. We vary from a constant, over a linear to a quadratic phase in the spatial pump distribution. The exact parameters for the simulation are given in \ref{app:numerical_simulation_of_the_waveguide_array}.

Note, that the absolute value of the $k^\perp$-correlations $|f(k_s^\perp,k_i^\perp)|^2$, as depicted in figure \ref{fig:phases}(b$_1$)-(b$_3$) remains unchanged, as there is no variation in the absolute value of the pump function. The phases of the $k^\perp$- correlations $\phi(f(k_s^\perp,k_i^\perp))$ however, vary with the induced phase, as depicted in figures \ref{fig:phases}(c$_1$)-(c$_3$).

In the case of a constant, zero-phase distribution in $\tilde{A}(k_s^\perp+k_i^\perp)$, the resulting real space correlations in figure \ref{fig:phases}(d$_1$) show the anti-bunching contribution, which has already been discussed in section \ref{sec:engineering_spatial_correlations_via_pump_shaping}.

Next, we introduce a linear phase in the pump shape, as depicted in figure \ref{fig:phases}(a$_2$). This is a trivial phase distribution, since a linear phase in \(k^\perp\)-space, can be realized by simply centering the spatial pump pattern around a waveguide different than zero. It leads to an asymmetry in the corresponding phase correlation as depicted in figure \ref{fig:phases}(c$_2$). Under Fourier transformation the phase asymmetry results, as expected, in a shift of the real space correlations, as given in figure \ref{fig:phases}(d$_2$), leaving the internal shape of the correlation function unchanged.

Quadratic phases, as shown in figure \ref{fig:phases}(a$_3$) introduce curvatures in the phases of the $k^\perp$-correlation (figure \ref{fig:phases}(c$_3$)). The result is a "stretching" of the real space correlations in figure \ref{fig:phases}(d$_3$). This is due to the linear slope of the quadratic functions, leading to different displacements in the real space correlations.

In general the shapes shown in figures \ref{fig:reshaping_spatial_spectral} and \ref{fig:phases} highlight both the flexibility and the limitations of PDC in WGA structures. More complicated pump shapes, like sinusodial shapes in the absolute value of the pump functions or different phase distributions only lead to various superpositions, displacements or stretching of the already discussed real space correlations in section \ref{sec:engineering_spatial_correlations_via_pump_shaping}.

\subsection{Enhanced engineering of spatial correlations via frequency filtering}\label{sec:enahnced_source_engineering}

We are able to further engineer the spatial correlations between the emitted photon pairs via frequency filtering \cite{branczyk_optimized_2010}. In this approach we use two distinct rectangular filters, with upper \(\omega_{s_{max}}, \omega_{i_{max}}\) and lower bounds \(\omega_{s_{min}}, \omega_{i_{min}}\), to select photon pairs within a narrow frequency range. This enables us to select a single point in the \( (\omega_s, \omega_i) \)-space depicted in figure \ref{fig:phase-mismatch}(a) and consequently the corresponding contour \(\Delta \beta_A\) in figure \ref{fig:phase-mismatch}(b)-(d). Post-selecting on both photons passing their individual filters, this procedure modifies the output state to
\begin{eqnarray}
    \fl \qquad \ket{\psi} = \frac{1}{\sqrt{\mathcal{N'}}} \hspace{-3mm} \int \limits_{\omega_{s_{min}}}^{\omega_{s_{max}}} \hspace{-3mm} \rmd\omega_{s} \hspace{-3mm} \int \limits_{\omega_{i_{min}}}^{\omega_{i_{max}}}  \hspace{-1mm} \rmd\omega_{i}  \int \limits_{-\pi}^{\pi} \hspace{-1mm} \rmd k^\perp_{s} \hspace{-2mm} \int \limits_{-\pi}^{\pi}  \hspace{-1mm} \rmd k^\perp_{i} \, f(\omega_s, \omega_{i}, k_s^\perp,k_i^\perp)  \hat{a}^\dagger(\omega_s,k_s^\perp)\hat{a}^\dagger(\omega_i,k_i^\perp) |0\rangle.
\label{eq:pdc_state_filtered}
\end{eqnarray}

\begin{figure}[htb]
    \begin{center}
        \includegraphics[width=0.9\textwidth]{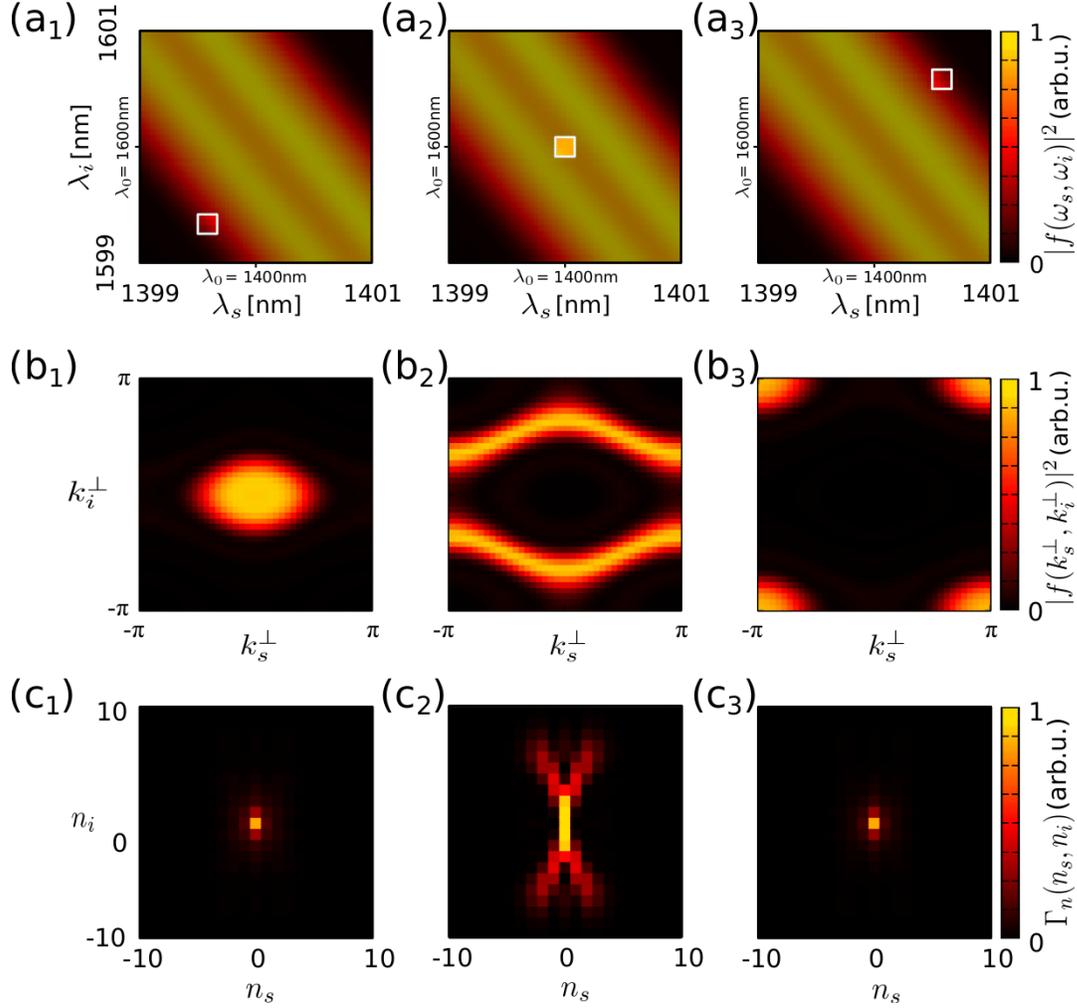}
    \end{center}
    \caption{Filtering specific wavelengths \( (\lambda_s, \lambda_i) \) from the PDC spectrum as indicated by the squares in (a\(_1\))-(a\(_3\)) selects different features in \(k^\perp\)-space (b\(_1\))-(b\(_3\)). The corresponding correlations in real space are depicted in (c\(_1\))-(c\(_3\)), where \(n\) labels the waveguide channel.}
    \label{fig:pump_omega_wave_dep}
\end{figure}

A particular example is visualized in figure \ref{fig:pump_omega_wave_dep}, for non-degenerate photon-pair filtering, which corresponds to figure \ref{fig:phase-mismatch}(c). In this scenario we assume signal photons with a wavelength of about 1400 nm and idler photons at about 1600 nm. The exact simulation parameters are given in \ref{app:numerical_simulation_of_the_waveguide_array}, while the corresponding coupling parameters are depicted in figure \ref{fig:C_lambda}. Figures \ref{fig:pump_omega_wave_dep}(a\(_1\))-(a\(_3\)), show the spectral shape of \(f(\omega_s, \omega_i, k_s^\perp, k_i^\perp)\), where the rectangle depicts the individual filter placement. Below, in figures \ref{fig:pump_omega_wave_dep}(b\(_1\))-(b\(_3\)) we depict the selected \( (k_s^\perp, k_i^\perp) \) correlations. Finally figures \ref{fig:pump_omega_wave_dep}(c\(_1\))-(c\(_3\)) show the corresponding spatial correlations \(\Gamma_n(n_s, n_i)\) (as defined in \ref{app:appendix_correlation_function}), obtained via Fourier transformation, where \( (n_s, n_i) \) label the individual waveguide channels.

\subsection{Summary: Spatio-spectral correlations}
In summary,  we explored the spatio-spectral structure of PDC in WGAs with special attention to spectrally dependent coupling effects. Our investigations revealed that the spatio-spectral correlations are fundamentally determined by the phase-mismatch \(\Delta \beta = \Delta \beta_\omega + \Delta \beta_A\). Furthermore the explicit frequency dependence of the coupling parameter \(C(\omega)\) leads to a spectral dependence of the emerging \(k^\perp\)-space correlations. Via pump shaping and adapted spectral filtering this system allows for the flexible preparation of photon pairs featuring a variety of spatial correlations. It is, however, fundamentally limited to correlations already present in the phase-mismatch \(\Delta \beta\).

\section{Experiment}\label{sec:experiment}
In this section we present a first preliminary experiment to test our theory and investigate the PDC process in a WGA. We examine the spatial and spectral properties of the PDC emission to reveal spectrally dependent coupling effects.

\begin{figure}[!htb]
    \begin{center}
        \includegraphics[width=0.9\textwidth]{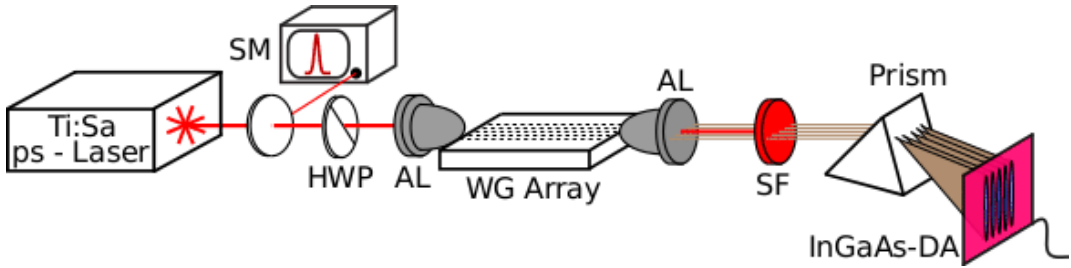}
    \end{center}
    \caption{Schematical picture of the experimental setup for measuring the spectral and spatial distributions of the PDC photons created in a PPLN WGA. Abbreviations: SM = spectrometer; HWP = half-wave plate;  AL = aspheric lens; WG = waveguide; SF = spectral filter; InGaAs-DA = InGaAs Detector array. For more details see text.}
    \label{fig:setup}
\end{figure}

In our experiment, schematically depicted in figure \ref{fig:setup}, we employed tunable picosecond Ti:Sapphire laser pulses (\unit{76}{\mega\hertz} repetition rate, \unit{775}{\nano\meter} central wavelength and \unit{0.8}{\nano\meter} bandwidth) as a pump for the PDC process in a \unit{40}{\milli\meter} long, type-I, PPLN WGA. The array  includes 101 waveguides packed into a region with a total width of \unit{1.6}{\milli\meter} and it is held during the experiment in an oven at the temperature of $185.0^{\circ}\mathrm{C}\pm 0.1^{\circ}\mathrm{C}$. A small portion of the pump beam is directed to a bright light spectrometer in order to control the pump wavelength. After passing through the power and polarization control (not shown), the pump beam illuminates a single waveguide. The PDC light generated in the pumped waveguide then couples to the neighbouring channels. The light  launched out of the WGA  was sent via a periscope (not shown) to a spectral filter, which blocks the residual pump. Thereafter, the spatially spread PDC emission passed through a dispersive prism and was imaged with a single lens (not shown) on an InGaAs-
detector array sensitive down to the few-photon level.
Note that by using a two-dimensional detector array, this configuration allows us to simultaneously resolve the properties of impinging light both spectrally and spatially. Thus, we can analyze any spatio-spectral correlations of the generated photons and compare it with our theoretical predictions.
The applied detector has a nearly flat response in the near infrared wavelengths. However, its sensitivity  drops fast in the vicinity of \unit{1.7}{\micro\meter}. The spectral resolution of our detector is approximately \unit{10}{\nano\meter}, which was calibrated with a tunable telecommunication continuous-wave laser coupled through a single isolated test waveguide on the same chip. This finite resolution introduces a smoothing of the experimental data.

\begin{figure}[!htb]
    \begin{center}
        \includegraphics[width=.9\textwidth]{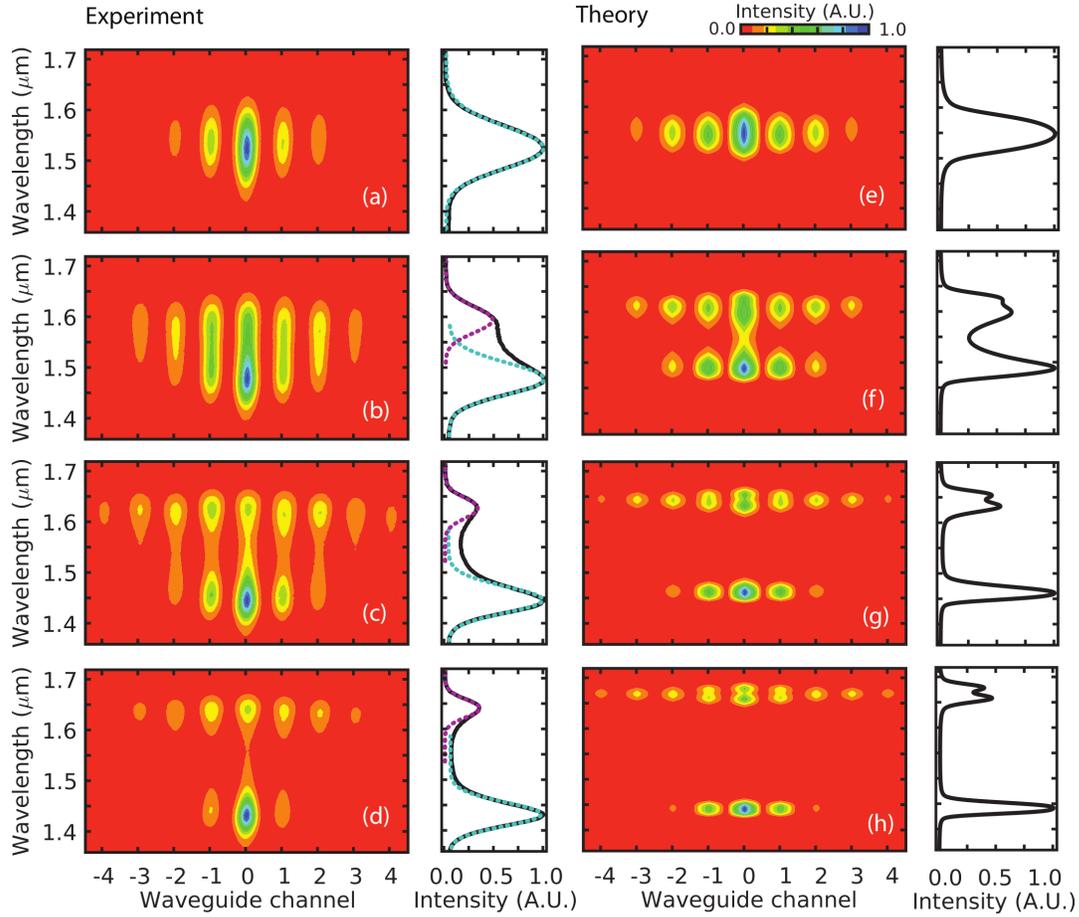}
    \end{center}
    \caption{Spectrally and spatially resolved PDC emission from the WGA. The contour plots illustrate the PDC light in different waveguide channels with respect to the wavelength. The spectra on right hand side show the marginal distribution of the central waveguide channel.
The measured data was recorded at the pump wavelengths near (a) \unit{774.9}{\nano\meter} (b) \unit{774.5}{\nano\meter} (c) \unit{774.2}{\nano\meter} and (d) \unit{773.9}{\nano\meter}. The numerical simulations in (e)-(h) correspond to  the measurements in (a)-(d), respectively. The solid lines illustrate the marginals, whereas the dotted lines are fits to the data to obtain the maxima of the spectral distributions.}
    \label{fig:results_coupling}
\end{figure}

After setting the pump wavelength close to the degeneracy point,  which was found via a measurement of the second-harmonic response in the WGA, we recorded images of the spatially and spectrally resolved PDC emission. Our results in figure \ref{fig:results_coupling}(a)-(d) show the measured signal and idler wavelengths in each waveguide channel, illustrating the spatio-spectral structure of the PDC state.

Close to the degeneracy point (figure \ref{fig:results_coupling}(a)), the spatial spread  spans over five waveguide channels  and the central channel has a spectral bandwidth of approximately \unit{100}{\nano\meter}. We then tuned the pump wavelength in small steps  from the degeneracy point towards lower wavelengths. As shown in figures \ref{fig:results_coupling}(b)-(d), we observe that the PDC emission is  separated into two spectral regions due to the curvature of the phase-matching function. Additionally,
one can recognize that the spatial spread of the PDC photons clearly behaves differently at the upper and lower spectral branches.
 The coupling of the PDC photons into the neighbouring channels increases with the growing wavelength and causes an obvious and detectable imbalance to the measured distributions.

In order to evaluate the quality of our theoretical model in section \ref{sec:pdc_in_nonlinear_waveguide_arrays}, we compare the experimental results with numerical simulations. Since our model explicitly takes into account the wavelength-dependent coupling, it explains the differences in the observed spatial spreads at different wavelengths.  The parameters governing the wavelength dependent coupling (see \ref{app:wavelength_dependency_of_the_coupling_parameter}) are determined by comparing the spatial spread given by the theoretical model to the measured ones. Additionally, the order of magnitude of the coupling parameter was confirmed by illuminating a single waveguide with laser featuring a fine linewidth, tuning the light in the range \unit{1520-1600}{\nano\meter} and measuring the linear spread of the light in the WGA.

For the comparison between the experimental and the numerical results, we have adapted the wavelength dependent coupling parameter together with 
the degeneracy wavelength of the process to reproduce the measured spatial 
spreads and central wavelengths of the PDC photons as well as their 
spectral marginals. 
The obtained numerical simulations depicted in figures \ref{fig:results_coupling}(e)-(h) show a very good quantitative agreement with the experimental data.  For the details of the simulation see \ref{app:numerical_simulation_of_the_waveguide_array}. Besides the wavelength-dependent spatial spread, reflecting clear correlations between the spatial and spectral degree of freedom, the theoretical model also predicts a spectral double peak structure for the non-degenerate PDC emission, which is clearly visible at the higher wavelength branch in figures \ref{fig:results_coupling}(f)-(h). This effect is caused by the splitting of the phase-matching function into two regions due to the different coupling parameters for signal and idler outside the degeneracy. We, however, are not able to observe this effect in the experiment due to the limited spectral resolution of the applied spectrometer. 
For the same reason the measured spectral marginals are slightly broader than the ones predicted with our theoretical model.

Due to the high pump powers utilized in our experiment, we cannot eliminate the possibility of higher order photon-number contributions. However, in our system, the generated photons are widely spread in the spatial and spectral domain, preventing stimulation or other high gain regime effects \cite{branczyk_non-classical_2010, christ_theory_2013}. In this regime, the generated higher order photon pairs are independent and possess the spatio-spectral characteristics, as discussed in section \ref{sec:spatio-spectral_dynamics_in_waveguide_arrays}. As such, our measurements are equivalent to measuring many photon-pair states subsequently.

\begin{figure}[!htb]
    \begin{center}
        \includegraphics[width=0.65\textwidth]{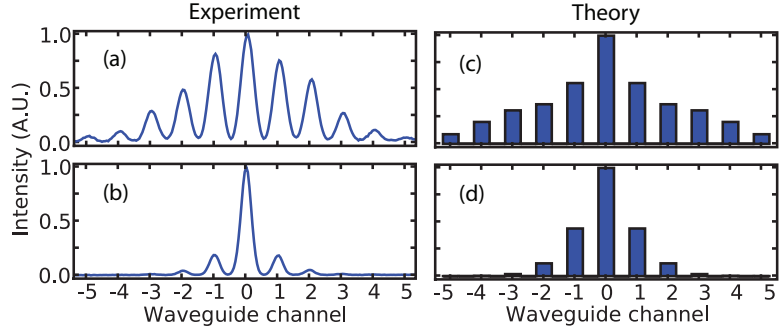}
    \end{center}
    \caption{The spatial marginal distributions deduced from figures \ref{fig:results_coupling} (d) and (h) for the upper and lower spectral branches.  The measured spatial marginals are illustrated at the wavelength regions of (a) \unit{1550-1750}{\nano\meter} and (b) \unit{1350-1550}{\nano\meter}. 
    The simulated spatial marginals are shown in (c) and (d) for the same spectral regions as in (a) and (b), respectively. The intensity in the central waveguide channel is in each case normalized 
to unity.}
    \label{fig:results_spatial}
\end{figure}

\begin{figure}[!htb]
    \begin{center}
        \includegraphics[width=0.55\textwidth]{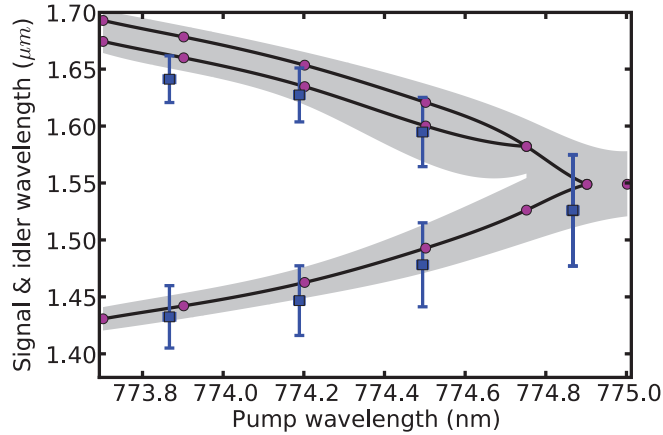}
    \end{center}
    \caption{The phase-matching curve of the WGA (squares) extracted from the marginal spectra in figure \ref{fig:results_coupling}(a)-(d). The error bars in the signal and idler wavelength are given by the full width at half maximum (FWHM) of the measured marginal distributions. Circles correspond to the simulated values including the  data shown in figure \ref{fig:results_coupling}(e)-(h), whereas the black solid lines and the shaded gray area  provide guides for the eye and represent the maxima and FWHM of the theoretical marginal spectra, respectively. As an effect of the wavelength dependent coupling, the theory predicts that the upper wavelength branch splits in two spectral regions.}
    \label{fig:results_pm}
\end{figure}

For a more direct test of our theoretical model with the experiment, we firstly compare in figure \ref{fig:results_spatial} the spatial marginal distributions from figures \ref{fig:results_coupling} (d) and (h)  both at the upper and the lower spectral branches. Although some differences appear in the shape of the spatial spreads between the experiment and theory, our model reproduces their widths correctly. Further, the measured spatial distributions also illustrate an excellent symmetry with respect to the central waveguide channel, which indicates a homogenous WGA quality.
Secondly, we 
deduce the phase-matching curve from the maxima of the measured spectral distributions in figure \ref{fig:results_coupling} and compare the experimental curve to the one gained from our theoretical model. As seen in figure \ref{fig:results_pm} our simulation accurately predicts the observed shape of the phase-matching curve. Small discrepancies arise from the accuracy at which the degeneracy point can be determined, limited resolution of the home-made spectrometer and fast decrease of its sensitivity at higher wavelengths. The splitting of the signal and idler spectra into two non-overlapping spectral regions caused by the curvature of the phase-matching function is already visible when tuning the pump wavelength by only a fraction of a nanometer from the degeneracy.


In regard to the spatio-spectral characteristics, we find a very good agreement between the theory for photon pairs and our experiment, even though our detector lacks true single-photon resolution and several photon pairs can be created by the pump at high power levels. Our results show that a detailed model, taking into account wavelength dependent coupling effects, is necessary in order to understand the experimental results. With our results we are  able to determine the relevant physical parameters required for measuring the desired correlation properties between the signal and idler photons.

\section{Conclusion}\label{sec:conclusion}

In this paper we have theoretically and experimentally investigated the PDC in nonlinear WGAs, explicitly taking into account spectrally dependent coupling effects. Our analysis revealed that the spatial and spectral degrees of freedom of the PDC process in the WGA are connected via the phase-mismatch \(\Delta \beta\). A phase-mismatch introduced in the spectral domain \(\Delta \beta_\omega\) is compensated via an opposite phase-mismatch in the spatial domain \(-\Delta \beta_A\), leading to spatio-spectral correlations in the generated quantum states. Thus, being able to select only specific regions of the phase-matching, by means of appropriate spectral filtering or by choosing the suitable spectral and spatial characteristics of the pump beam, we can modify the joint spatio-spectral amplitude. Consequently, we are able to engineer the emerging spatial correlations of the two-photon state. This leads to a variety of well defined spatial correlation patterns between the signal and idler photons limited only by the dispersion relation of the transverse momentum.

Our experimental investigations show that a wavelength dependent coupling parameter, which quantifies the strength of the evanescent overlap within the WGA, must be taken into account to understand the properties of the generated two-photon states. With our model, which takes into account the wavelength dependency of the coupling constant, we were able to accurately predict the spatial spread of the PDC photons into the adjacent waveguide channels at different wavelength regimes. Furthermore, we found that the phase-matching curve of the PDC process, extracted from the measured data is in very good agreement with the theoretical prediction.

In conclusion, our analysis provides a framework for a detailed understanding of the possibilities and limitations of quantum state generation in nonlinear WGAs and enables an accurate engineering of a variety of spatio-spectral correlations. Our results are important for the advancement of future applications such as quantum walks or other sophisticated optical networks on WGAs.

\section{Acknowledgments}
The authors thank Harald Herrmann and Hubertus Suche for useful discussions and helpful comments. The authors further thank Frank Setzpfandt and Thomas Pertsch for discussions concerning the experimental setup.

A.~G., C.~S.~H.\ and I.~J.\ received funding from MSM 6840770039, RVO 68407700 and GACR 13-33906S. A.~G.\ acknowledges partial support from the National Research Fund of Hungary under contract No. T83858.

\begin{appendix}

\section{Wavelength dependency of the coupling parameter}
\label{app:wavelength_dependency_of_the_coupling_parameter}

\begin{figure}[htb]
    \begin{center}
        \includegraphics[width = 0.5\textwidth]{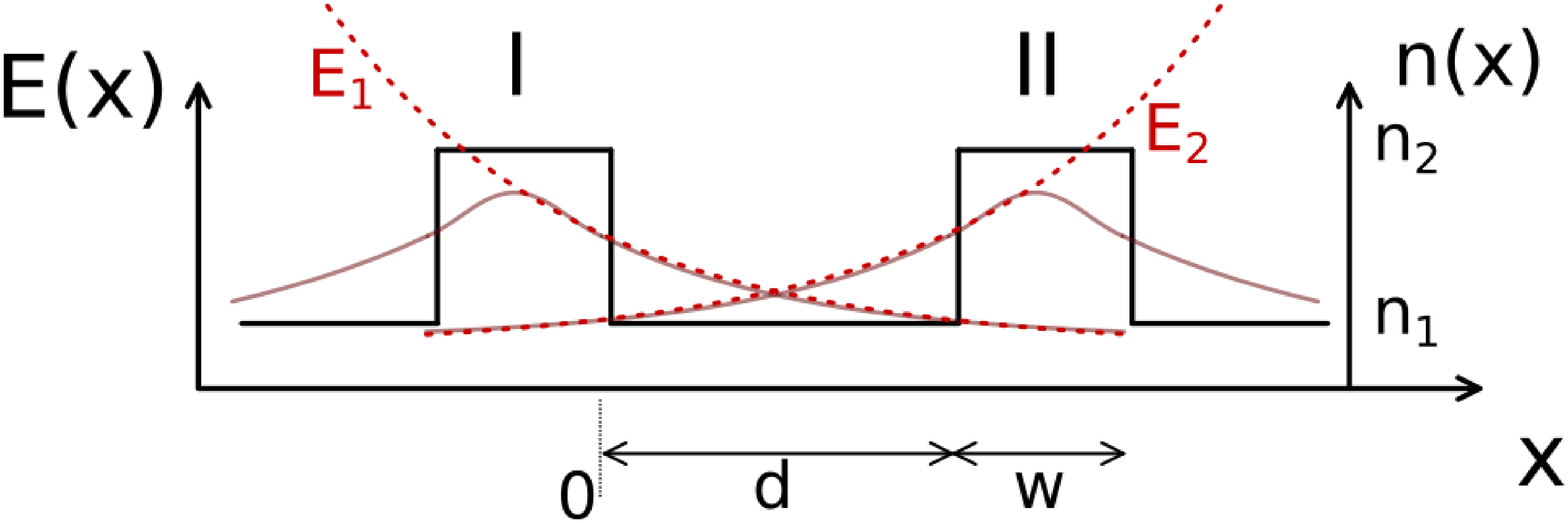}
    \end{center}
    \caption{Form of the electric fields for a rectangular refractive index profile. The overlap of the electric fields determines the coupling parameter.}
    \label{fig:coupling_model}
\end{figure}

Only over a narrow frequency range the coupling parameter \(C(\omega)\) can be approximated as a constant \(C_0\) \cite{lederer_discrete_2008, solntsev_spontaneous_2012}. In this paper, however, we consider spectral widths spanning over 100 nm in range. We therefore include wavelength dependent coupling effects \(C(\omega)\) into our analysis \cite{szameit_hexagonal_2006}. Much work on the frequency dependence of \(C(\omega)\) in LiNbO$_3$ has already been performed \cite{alferness_characteristics_1979, ctyroky_3-d_1984}, but it has mainly relied on numerical calculations. Here we consider a simple model for the wavelength dependence of the coupling parameter \(C(\omega)\), as shown in figure \ref{fig:coupling_model}.

In coupled mode theory, the coupling parameter $C(\omega)$ is given by the overlap integral from the mode of waveguide (I) with the mode in the neighbouring waveguide (II) as described in \cite{yariv_quantum_1989}
\begin{eqnarray}
C(\lambda)= \mathcal{C}_0 \frac{2\pi}{\lambda} \int \limits_{d}^{d+w} \Delta n^2(x) E_1^{*}(x)E_2(x) \, \mathrm d x.
\label{eq:Cparameter}
\end{eqnarray}
Here $\mathcal{C}_0$ is a constant depending on the mode number and $\Delta n(x)$ is the refractive index difference between the waveguide \(n_2\) and the surrounding bulk material \(n_1\) \cite{ctyroky_3-d_1984} and is zero outside waveguides (I) and (II).
Assuming a very simple model for the waveguide modes, which only considers the exponential part of the evanescent field, we arrive at
\begin{eqnarray}
E_1(x)&=E_{1_0} e^{-\gamma x}\\
E_2(x)&=E_{2_0} e^{\gamma(x-d)}.
\end{eqnarray}
Here $E_{i_0}$ are the amplitudes of the electric field, and $\gamma$ is the damping factor determined by the difference between the propagation vector in the waveguide $\beta^{(0)}$ and in bulk material $k$ \cite{tamir_integrated_1979}. We model the wavelength dependence of this parameter via $\gamma \propto \frac{n(\lambda)}{\lambda}$, where \(n(\lambda)\) describes the refractive index in terms of wavelength $\lambda$. We further neglect the possible influence of the waveguide dopant on the wavelength dependent refractive index. Solving (\ref{eq:Cparameter}) with these assumptions and integrating over the region of waveguide (II) yields
\begin{eqnarray}
C(\lambda)= \mathcal{C} \frac{1}{\lambda} \exp\left(-\gamma_0\frac{n(\lambda)}{\lambda}\right),
\end{eqnarray}
where $\mathcal{C}$ is a constant depending on the width of the waveguides and their mode profiles, as well as on the refractive index difference between the waveguide and substrate and $\gamma_0$ is a constant that is determined by the distance between the waveguides and the waveguide mode profiles.

\section{Spatio-spectral correlation function}
\label{app:appendix_correlation_function}

The most straightforward way to obtain information about the joint spatio-spectral distribution of signal and idler are spectrally and/or spatially resolved correlation function measurements. The correlation function between one photon at \( (\omega_s, k_s^\perp)\) and the other photon at \( (\omega_i, k_i^\perp)\) from the two-photon PDC state in \eref{eq:pdc_state} is given as:
\begin{eqnarray}
\nonumber
\fl \qquad \tilde{\Gamma}_{k,\omega}(k^{\perp}_{s}, k^{\perp}_{i}, \omega_{s}, \omega_{i})& = & \braket{  {\psi} |\hat{a}^{\dagger}(\omega_s, k_s^{\perp})  \hat{a}^{\dagger}(\omega_i, k_i^{\perp})  \hat{a}(\omega_i, k_i^{\perp})  \hat{a}(\omega_s, k_s^{\perp}) | {\psi}} \\
& = & \left\{
      \begin{array}{l l}
        4/\mathcal{N} \big | f (\omega_s, \omega_{s}, k^{ \perp}_s ,k^{ \perp}_s)\big|^{2}& \quad \textrm{if $\omega_s = \omega_i$ and $k_s^\perp = k_i^\perp$}\\
        1/\mathcal{N} \big | f (\omega_s, \omega_{i}, k^{ \perp}_s ,k^{ \perp}_i)\big|^{2} & \quad \textrm{else}.
      \end{array} \right.
\end{eqnarray}
By tracing over the spectral degree of freedom, the correlation function in $k^\perp$-space can be written in the form
\begin{eqnarray}
\tilde{\Gamma}_k(k^{\perp}_{s}, k^{\perp}_{i} ) = \int d \omega_{s} \int d \omega_{i} \ \tilde{\Gamma}_{k,\omega}(k^{\perp}_{s}, k^{\perp}_{i}, \omega_{s}, \omega_{i}).
\end{eqnarray}
In the real space the correlation function is determined as
\begin{eqnarray}
\Gamma_n(n_{s}, n_{i}) = \int d \omega_{s} \int d \omega_{i} \ \Gamma_{n,\omega}(n_{s}, n_{i}, \omega_{s}, \omega_{i}),
\end{eqnarray}
in which  $\Gamma_{n,\omega}(n_{s}, n_{i}, \omega_{s}, \omega_{i}) = \braket{  {\Psi} |\hat{a}^{\dagger}(\omega_s, n_s)  \hat{a}^{\dagger}(\omega_i, n_i)  \hat{a}(\omega_i, n_i)  \hat{a}(\omega_s, n_s) | {\Psi}}$, and the bi-photon state $\ket{\Psi}$ in real space, is given by a two-dimensional Fourier transformation of $\ket{{\psi}}$ in $k^\perp$-space.

\section{Numerical simulation of the WGA}
\label{app:numerical_simulation_of_the_waveguide_array}

In the scope of this paper we consider extraordinarily polarized type-I PDC in lithium niobate using the dispersion relations from \cite{jundt_temperature-dependent_1997}. We run our experiments far above room temperatures. Consequently our theoretical model takes into account the temperature dependence of the refractive indices, also given in \cite{jundt_temperature-dependent_1997}.

We adjust the signal and idler frequencies with the help of quasi phase-matching \cite{christ_spatial_2009}, where a periodic sign change of the $\chi^{(2)}$ nonlinearity in the material introduces a modification in the spectral phase-matching condition
\begin{eqnarray}
\Delta \beta_{\omega}(\omega_{s}, \omega_{i}) = \beta^{(0)}_p(\omega_s+\omega_{i})-\beta^{(0)}(\omega_s)-\beta^{(0)}(\omega_i)-\frac{2\pi}{\Lambda_{\textrm{eff}}}.
\end{eqnarray}
In the simulation $\Lambda_{\textrm{eff}}$ is an effective grating period, fitting the theoretically predicted wavelengths of the phase-matched PDC process to the experimental values.

\begin{figure}[htb]
    \begin{center}
        \includegraphics[width = 0.5\textwidth]{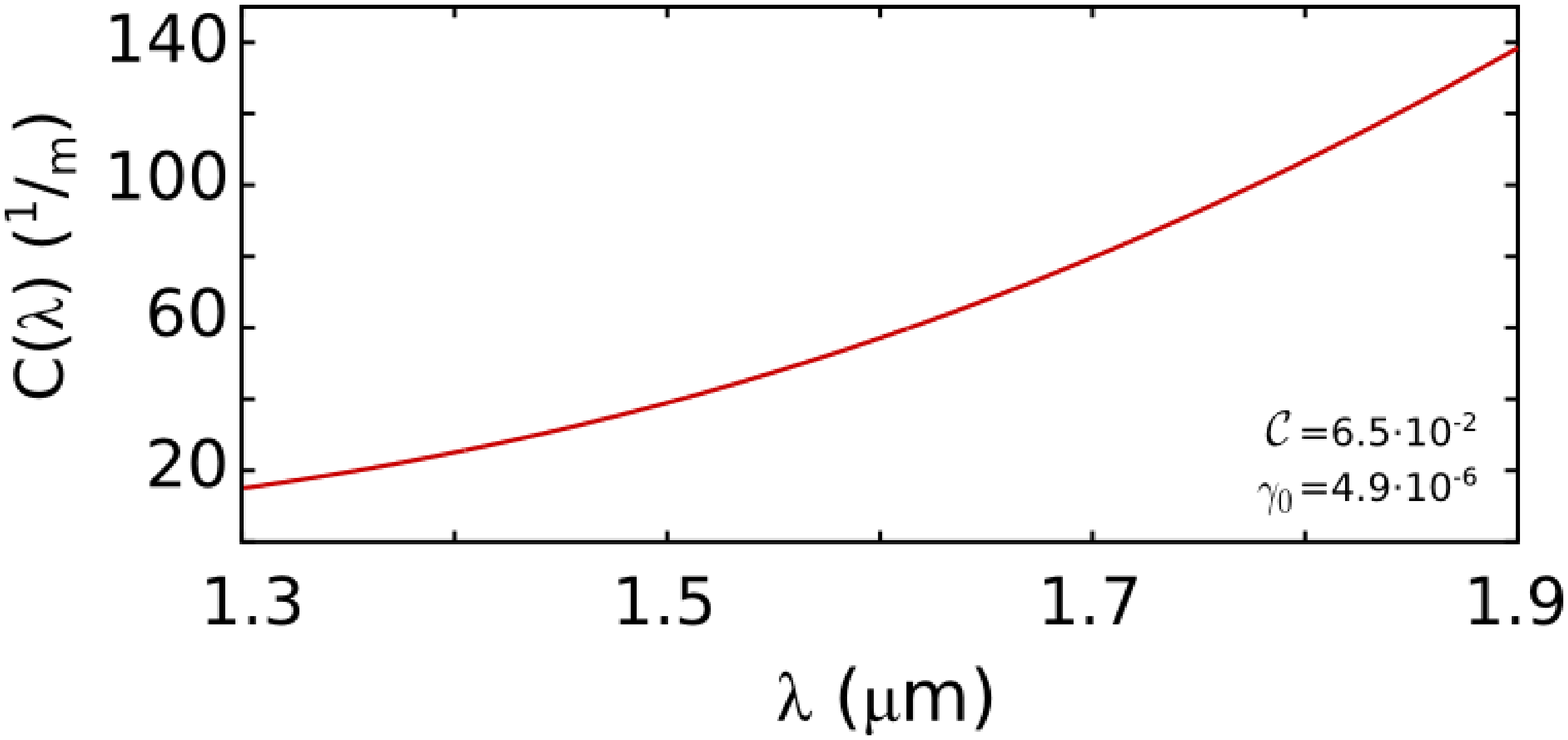}
    \end{center}
    \caption{Wavelength dependence of the coupling parameter \(C(\lambda)\) used to fit the numerical simulations to the experimental results.}
    \label{fig:C_lambda}
\end{figure}

The wavelength dependent coupling parameter \(C(\lambda)\) is fitted to the experimental results using the description in \ref{app:wavelength_dependency_of_the_coupling_parameter}. Its exact form is depicted in figure \ref{fig:C_lambda}.

The process properties for the numerical modelling are given in table \ref{tab:numerical_parameters}, while the experimental parameters can be found in section \ref{sec:experiment}.

\begin{table}[htb]
\begin{center}
\begin{tabular}{ll}
\hline
\multicolumn{2}{c}{Properties of WGA (numerical simulation)}\\
\hline
Waveguide length $L$ [\hspace{-0.5ex}\unit{}{\metre}] & 0.04 \\
Total number of waveguides on the array & 41 \\
Temperature $T$ [\hspace{-0.5ex}\unit{}{\celsius}] & 185.0 \\
Effective coupling parameter $\mathcal{C} $ & $6.5\cdot10^{-2}$\\
Effective damping coefficient $\gamma_0$ [m] & $4.9\cdot 10^{-6}$\\
\hline
\multicolumn{2}{c}{Phase engineering (Figure \ref{fig:phases})}\\
\hline
Phase-matched wavelength pair ($\lambda_s$, $\lambda_i$) [nm] & (1550.0,1550.0)\\
Narrowband filtering ($\lambda_{min}$, $\lambda_{max}$) [nm] & (1549.9, 1550.1)\\
Real valued spatial pump width $dk_p^\perp$ & $\frac{\pi}{2}$\\
Number of waveguides in the array (only this section) & 49\\
Effective coupling parameter $\mathcal{C} $ & $13\cdot10^{-2}$\\
Effective damping coefficient $\gamma_0$ [m] & $4.9\cdot 10^{-6}$\\
\hline
\multicolumn{2}{c}{Correlation calculation (Figure \ref{fig:pump_omega_wave_dep})}\\
\hline
Phase-matched wavelength pair ($\lambda_s$, $\lambda_i$) [nm] & (1400.0,1600.0)\\
Effective coupling parameter $\mathcal{C} $ & $13\cdot10^{-2}$\\
Effective damping coefficient $\gamma_0$ [m] & $4.9\cdot 10^{-6}$\\
\hline
\multicolumn{2}{c}{Experimental comparison (Figure \ref{fig:results_coupling}(e-h))}\\
\hline
Phase-matched wavelength pair ($\lambda_s$, $\lambda_i$) [nm] & (1549.8,1549.8)\\
Spectral pump bandwidth $2\pi\Delta \lambda_p$  [nm] & 0.5\\
\hline
\multicolumn{2}{c}{Near-degenerate correlation calculation (Figure \ref{fig:spectral_shaping_sim})}\\
\hline
Phase-matched wavelength pair ($\lambda_s$, $\lambda_i$) [nm] & (1550.1,1550.1)\\
Spectral pump bandwidth $\Delta \lambda_p$  [nm] & 0.5\\
Coupling parameter $C_0$ [1/m] & 400\\
Number of waveguides in the array (only this section) & 51\\
\hline
\end{tabular}
\end{center}
\caption{Parameters for the simulation of the spatial correlations in figure \ref{fig:pump_omega_wave_dep} and the modelling of the experimental results in figure \ref{fig:results_coupling} and \ref{fig:results_pm}.}
\label{tab:numerical_parameters}
\end{table}

\begin{figure}
\begin{center}
\includegraphics[width=1.\textwidth]{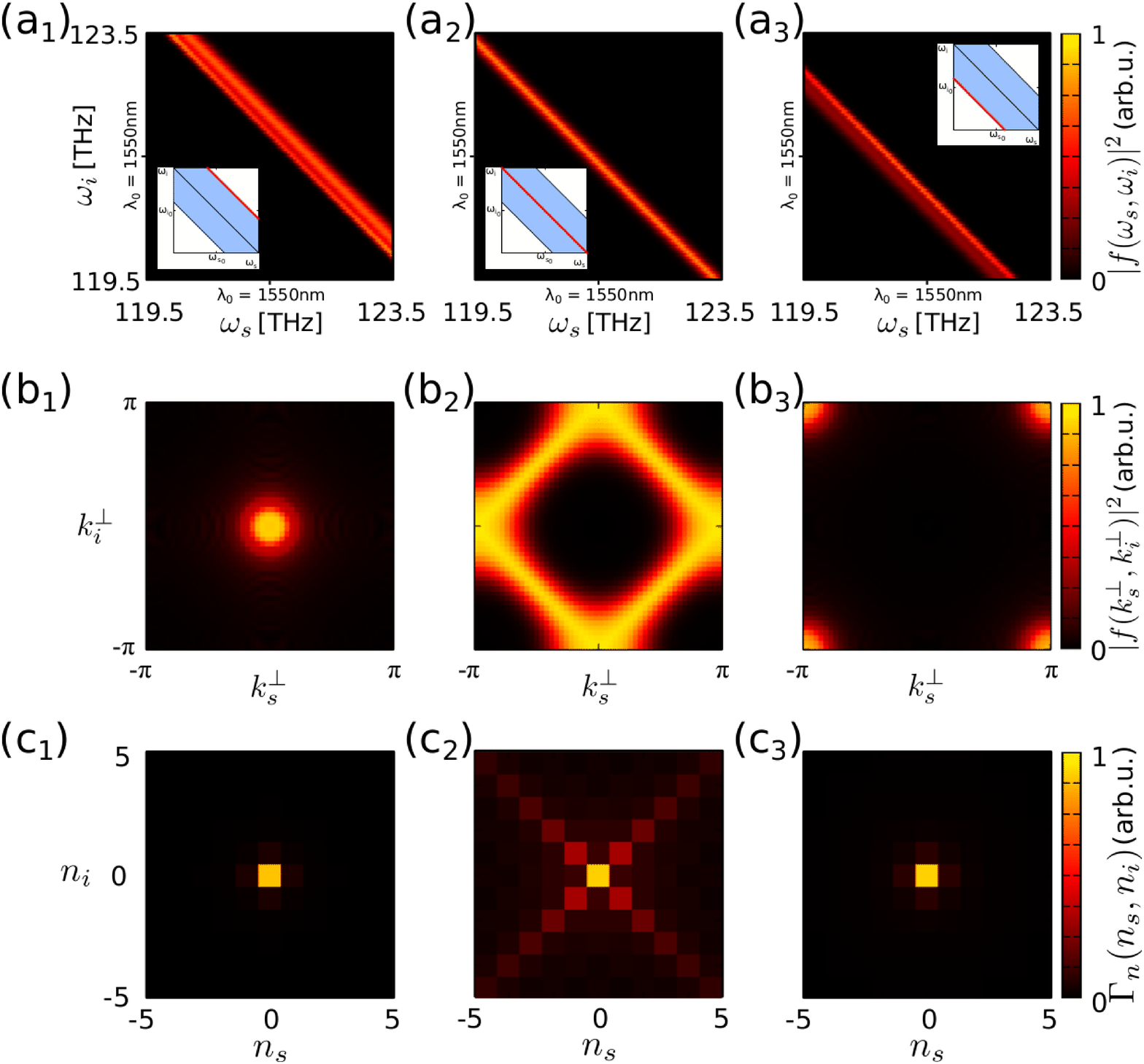}
\end{center}
\caption{Numerical simulation of the output correlation patterns in the WGA by exploiting the tunability of the correlations via the spectral pump shape. (a$_1$)-(a$_3$) show the corresponding joint spectral amplitudes of the process, illustrating the different pump frequencies. The inlays give the position of the pump shape in relation to the phase-matching function. The resulting correlation functions in \(k^\perp\)-space are depicted in figures (b$_1$)-(b$_3$), according to the discussed spatio-spectral interplay in figure \ref{fig:phase-mismatch}(a) and (b). In figures (c$_1$)-(c$_3$) the real space correlation function of the WGA is shown. We observe a distinct X-like shape in figure (c$_2$), while the coincidence intensity in figures (c$_1$) and (c$_3$) is localized around the pumped waveguide.}
\label{fig:spectral_shaping_sim}
\end{figure}

Figure \ref{fig:spectral_shaping_sim} shows the output correlations available when pumping near the degeneracy wavelength, corresponding to figure \ref{fig:phase-mismatch}(b) in section \ref{sec:spatio-spectral_dynamics_in_waveguide_arrays}. Here, we exploit the tunability of the spatial output correlation patterns via the spectral pump shape, as discussed in section \ref{sec:the_phasematching_function}.
 
The selection process is shown from top to bottom. In figures \ref{fig:spectral_shaping_sim}(a$_1$)-(a$_3$) we select the spatial correlations by tuning the pump frequency according to the selection rules discussed in section \ref{sec:the_phasematching_function}. The small inlays depict the position of the pump shape in relation to the phase-matching function, serving as orientation in the spectral domain. Figures \ref{fig:spectral_shaping_sim}(b$_1$)-(b$_3$) depict the selected spatial correlation in $k^\perp$-space in accordance to the spatio-spectral interplay shown in figure \ref{fig:phase-mismatch}(a) and (b). The corresponding real space correlation functions in figure \ref{fig:spectral_shaping_sim}(c$_1$)-(c$_3$) are then determined via Fourier transformation.
\clearpage

\end{appendix}
\newpage

\section*{References}
\bibliographystyle{unsrt}
\bibliography{WGArrays}

\end{document}